\documentstyle[12pt]{article}
\input{epsf}
\newcommand{\be}{\begin{eqnarray}}
\newcommand{\ee}{\end{eqnarray}}
\newcommand{\ba}{\begin{array}}
\newcommand{\ea}{\end{array}}
\newcommand{\half}{{\textstyle{\frac{1}{2}}}}
\newcommand{\textfrac}[2]{{\textstyle{\frac{#1}{#2}}}}
\newcommand{\fourint}[1]{\int\!\frac{d^4 #1}{(2\pi)^4}}
\newcommand{\tr}{{\rm tr}\,}
\newcommand{\nablaslash}{\nabla\hspace{-.65em}/\hspace{.3em}}

\newcommand{\kslash}{k\hspace{-.5em}/\hspace{.15em}}
\newcommand{\effop}[1]{\mbox{``$#1$''}} 
\begin{document}
%
%
\rightline{RUB-TPII-33/95}
\rightline{hep-ph/9607244}
\vspace{.3cm}
\begin{center}
\begin{large}
{\bf Mixed quark--gluon condensate from instantons} \\
\end{large}
\vspace{1.4cm}
{\bf M.V.\ Polyakov}$^{\rm 1, 2, 3}$
{\bf and C. Weiss}$^{\rm 4}$ \\
\vspace{0.2cm}
{\em Institut f\"ur Theoretische Physik II \\
Ruhr--Universit\"at Bochum \\
D--44780 Bochum, Germany}
\end{center}
\vspace{1cm}
\begin{abstract}
\noindent
We calculate the vacuum expectation value of the dimension--5 ``mixed'' 
quark--gluon operator 
$O_\sigma = \bar\psi \half \lambda^a \sigma_{\mu\nu} \psi F_{\mu\nu}^a$ 
in the instanton vacuum. Within the $1/N_c$--expansion the QCD operator 
is replaced by an effective many--fermion operator, which is averaged over 
the effective theory of massive quarks derived from instantons. We find 
$m_0^2 \equiv \langle O_\sigma \rangle / \langle \bar\psi\psi \rangle 
\approx 4 \bar\rho^{-2} = 1.4\, {\rm GeV}^2$, somewhat larger than
the estimate from QCD sum rules for the nucleon.
\end{abstract}
\vspace{1cm}
\vfill
\rule{5cm}{.15mm}
\\
\noindent
{\footnotesize $^{\rm 1}$ Permanent address: Petersburg Nuclear Physics
Institute, Gatchina, St.\ Petersburg 188350, Russia} \\
{\footnotesize $^{\rm 2}$ Supported by a fellowship of the 
A.v.Humboldt Foundation} \\
{\footnotesize $^{\rm 3}$ E-mail: maximp@hadron.tp2.ruhr-uni-bochum.de} \\
{\footnotesize $^{\rm 4}$ E-mail: weiss@hadron.tp2.ruhr-uni-bochum.de}
%
%
%
%
\newpage
The non-perturbative structure of the QCD vacuum can be expressed in
terms of non-zero vacuum expectation values of various composite operators.
These ``condensates'' have been introduced as phenomenological parameters 
in a non-perturbative generalization of the operator product 
expansion, which can be related to observable hadronic properties by 
the sum rule method \cite{SVZ79,RRY85,Ioffe81}. The lowest--dimensional
condensates, $\langle F_{\mu\nu}^2 \rangle$ and
$\langle \bar\psi \psi \rangle$, are
well-known from phenomenology. Considerable uncertainty
prevails about the values of the higher--dimensional
condensates. The four--quark condensates, $\langle 
\bar\psi \Gamma \psi \bar\psi \Gamma \psi \rangle$, are usually 
estimated assuming factorization, an approximation justified 
in the large--$N_c$ limit \cite{SVZ79}. No such simple estimate
exists for the condensate of the dimension--5
mixed quark--gluon operator,
\be
O_\sigma &=& \bar\psi \half \lambda^a \sigma_{\mu\nu} \, \psi 
F^a_{\mu\nu} ,
\label{O_sigma}
\ee
which enters in QCD sum rules for the nucleon 
channel \cite{Ioffe81}, for exotic light mesons \cite{BDY}
and for heavy--light mesons \cite{Zhit93}. The value of this
condensate is usually expressed as a multiple of the quark condensate,
\be
\langle O_\sigma \rangle &=& m_0^2 \langle \bar\psi \psi \rangle .
\label{m_0_def}
\ee
The parameter $m_0^2$ has been estimated by Belyaev and Ioffe from
sum rule stability \cite{BI82}. At a normalization point of 
$500\,{\rm MeV}$ they find $m_0^2 = (0.8 \pm 0.2)\, {\rm GeV}^2$.
\par
In a more microscopic picture of the QCD vacuum, the condensates
can be understood as being generated by certain non-perturbative 
fluctuations of the fields. In particular, instantons are
responsible for the spontaneous breaking of chiral symmetry. 
A quark condensate arises due the delocalization of the fermion 
zero modes associated with the individual instantons in the 
medium \cite{DP86}. Assuming the gluon condensate to be entirely
due to instantons results in a density of the instanton medium
which gives a consistent description not only of the quark condensate, 
but also of meson and baryon characteristics \cite{Sh82}. One may 
therefore assume that instantons are also the dominant non-perturbative
fluctuations giving rise to the mixed condensate, eq.(\ref{m_0_def}).
\par
In this letter, we evaluate the vacuum average of the
mixed operator, eq.(\ref{O_sigma}), on the basis of the quantitative 
theory of the instanton medium of Diakonov and Petrov \cite{DP84_1}.
The calculation of $\langle O_\sigma \rangle$ requires averaging over 
the gluon field, {\em i.e.}, the instanton coordinates, as well as over 
the quark field, taking into account the dynamical breaking of chiral symmetry.
Such averages can be computed using a recently developed method \cite{DPW95}, 
which makes use of the $1/N_c$--expansion. Integrating first
over the instanton coordinates, one derives from the instanton 
vacuum an effective quark action of the form of a Nambu--Jona-Lasinio 
model, which describes quarks with a (momentum--dependent)
dynamical mass \cite{DP86_prep}. In this effective theory QCD operators 
involving gluon fields can systematically be represented as many--fermion
operators, whose averages can be evaluated with standard techniques.
With this method we compute
$\langle O_\sigma \rangle$ at the scale 
at which the QCD coupling constant is defined in the instanton vacuum, 
typically of order of the inverse average instanton size, 
$\bar\rho^{-1} = 600\, {\rm MeV}$. We compare our result with 
the QCD sum rule estimate. Furthermore, we show that in the 
instanton vacuum the parameter $m_0^2$ of eq.(\ref{m_0_def}) has a 
simple interpretation as a ratio of one--instanton matrix elements. 
\par
We begin by rewriting the operator eq.(\ref{O_sigma}) in terms of 
euclidean gluon and quark fields
($\psi^\dagger \equiv i \bar\psi$),
\be
O_\sigma (x) &=& i \psi^\dagger (x) \half \lambda^a \sigma_{\mu\nu} \, 
\psi (x) F^a_{\mu\nu} (x) .
\label{O_sigma_eucl}
\ee
Following \cite{DPW95}, we replace the gluonic part of this operator, 
$F_{\mu\nu}^a (x)$, by an effective quark operator.  
For simplicity, we assume only one quark flavor in the following; the 
generalization to $N_f > 1$ is straightforward and will be discussed
below. 
\par
In zero mode approximation \cite{DP86}, the interaction of the quark field 
with an (anti--) instanton with collective coordinates $\rho$ (size), 
$z$ (center) and $U$ ($SU(N_c)$ matrix describing color orientation) 
is given by
\be
V_\pm [\psi^\dagger , \psi ] 
&=& 4\pi^2\rho^2
\fourint{k_1}\fourint{k_2} \exp \left( i z\cdot (k_2 - k_1 )\right) \;
F(k_1 ) F(k_2 ) \nonumber \\
&& \times \; \psi^\dagger_{i\gamma} (k_1 ) \;
\left[ \frac{1}{8}
\gamma_\kappa \gamma_\lambda \frac{1 \pm \gamma_5}{2} \right]_j^i \;
\left[ U \tau^\mp_\kappa \tau^\pm_\lambda U^\dagger \right]^\gamma_\delta \;
\psi^{j\delta} (k_2 ) .
\label{V_I}
\ee
Here $F(k)$ is a form factor of width $\rho^{-1}$,
proportional to the Fourier transform of the
wave function of the fermion zero mode, with $F(0) = 1$,
and $\tau^\pm_\kappa$ are $N_c \times N_c$ matrices with
$(\tau, \mp i)$ in the upper left corner and zero 
elsewhere \cite{DP86}. The field strength of the (anti--) 
instanton is given as a function of the collective coordinates, 
\be
F^a_{\pm\mu\nu} (x; z, U) &\equiv& 
\half \tr [\lambda^a U \lambda^b U^\dagger ] F^b_{\pm\mu\nu} (x - z) , 
\nonumber \\
U \left[ \half \lambda^b F_{\pm\mu\nu}^b (x - z) \; \right] U^\dagger
&=& \half \lambda^a F^a_{\pm\mu\nu} (x; z, U) , 
\label{F_U}
\ee
where $F_{\mu\nu}^b (x)$ denotes the instanton field (in singular gauge)
in standard orientation,
\be
F^b_{\pm\mu\nu}(x) &=& \frac{8\rho^2}{(x^2 + \rho^2 )^2}
\left[ (\eta^\mp )^b_{\rho\nu} \frac{x_\rho x_\mu}{x^2}
\; + \; (\eta^\mp )^b_{\mu\rho} \frac{x_\rho x_\nu}{x^2}
\; - \; \half (\eta^\mp )^b_{\mu\nu} \right] ,
\label{F_inst}
\ee
$(\eta^+ )^b_{\mu\nu} = \eta^b_{\mu\nu}, 
(\eta^- )^b_{\mu\nu} = \bar\eta^b_{\mu\nu}$ are the 't Hooft symbols.
The effective fermion operator corresponding to $F_{\pm\mu\nu}^a (x)$ is
defined as the average of the product of the instanton field, 
eq.(\ref{F_U}), with the instanton--quark interaction, 
eq.(\ref{V_I}), over the collective coordinates of one 
instanton \cite{DPW95},
\be
(Y_{F\pm})^a_{\mu\nu}(x)[\psi^\dagger , \psi ]
&\equiv & i \frac{N_c M}{4\pi^2\bar\rho^2} \int d^4 z \int dU \; 
F^a_{\pm\mu\nu} (x; z, U) \; V_\pm [\psi^\dagger , \psi ] 
\label{Y_1} \\
&=& i N_c M \int d^4 z \; F^b_{\mu\nu} (x - z) 
\fourint{k_1}\fourint{k_2} \exp \left( i z \cdot (k_2 - k_1 )\right)
\nonumber \\
&& \times F(k_1 ) F(k_2 ) 
(S_{\pm})_{\kappa\lambda}^{abc} \; \psi^\dagger_{i\alpha} (k_1 ) 
\; \left[\frac{1}{8} 
\gamma_\kappa \gamma_\lambda \frac{1 \pm \gamma_5}{2} \right]_j^i
\left(\frac{\lambda^c}{2}\right)^\alpha_\beta \; \psi^{j\beta} (k_2 ) .
\nonumber
\ee
The normalization factor of the effective operator is proportional
to the dynamically generated quark mass at zero euclidean 
momentum, $M$. When integrating over instanton sizes 
we have assumed all instantons to be of the same size, $\bar\rho$, 
which is justified by the fact that the width of the effective
size distribution of instantons in the medium 
is of order $1/N_c$ \cite{DP84_1,DPW95}. By 
$(S_{\pm})_{\kappa\lambda}^{abc}$ we denote the average over color 
orientations with the Haar measure of $SU(N_c )$, 
\be
(S_{\pm})_{\kappa\lambda}^{abc} &=& 
\half (\lambda^a )^\beta_\alpha \, (\lambda^b )^{\alpha'}_{\beta'} \,
(\lambda^c )^\delta_\gamma \,
(\tau^\mp_\kappa \tau^\pm_\lambda )^{\gamma'}_{\delta'}
\int dU \, U^\alpha_{\alpha'} \, U^\gamma_{\gamma'}
(U^\dagger )^{\beta'}_\beta \, (U^\dagger )^{\delta'}_\delta 
\nonumber \\
&=& \frac{2}{N_c^2} \, \delta^{ac} \,
i (\eta^\mp )^b_{\kappa\lambda} ,
\ee
where the last equality is understood to leading order in $1/N_c$. 
The spin structure of the fermion vertex in eq.(\ref{Y_1}) is
simplified using the 
identity\footnote{To avoid confusion we note that we are using the 
convention $\gamma_5 = \gamma_1 \gamma_2 \gamma_3 \gamma_4$
for the euclidean $\gamma_5$, as in \cite{DP86}.}
\be
i (\eta^\mp )^b_{\mu\nu} (\eta^\mp )^b_{\kappa\lambda} \, 
\gamma_\kappa \gamma_\lambda &=&
(\eta^\mp )^b_{\mu\nu} (\eta^\mp )^b_{\kappa\lambda} \, 
\sigma_{\kappa\lambda} \, 
\;\; = \;\; 4 \sigma_{\mu\nu} \frac{1 \pm \gamma_5}{2} .
\label{eta_eta}
\ee
Furthermore, we introduce the Fourier transform of the instanton 
field strength,
\be
\int d^4 x \, F^a_{\pm\mu\nu} (x) \exp (-ik\cdot x)
&=& \rho^2 G(k) 
\left[ (\eta^\mp )^a_{\rho\nu} \frac{k_\rho k_\mu}{k^2}
\; + \; (\eta^\mp )^a_{\mu\rho} \frac{k_\rho k_\nu}{k^2}
\; - \; \half (\eta^\mp )^a_{\mu\nu} \right] , 
\hspace{.5cm}
\label{F_fourier} 
\ee
\be
G(k) &=& 8 \int d^4 x \frac{1}{(x^2 + \rho^2 )^2}
\left[ 1 + \frac{4}{3} \left(\frac{(k\cdot x)^2}{k^2 x^2} - 1\right)
\right] \exp (-ik\cdot x) 
\nonumber \\
&=& 32 \pi^2 \left\{ \frac{1}{2} K_0 (t)
+ \left[ \frac{4}{t^2} K_0 (t) + \left( \frac{2}{t} + \frac{8}{t^3}
\right) K_1 (t) - \frac{8}{t^4} \right]
\right\}, \hspace{1cm} t \; = \; k\rho , 
\label{G} 
\ee
where $K_n (t)$ are modified Bessel functions of the second kind.
Note that the tensor structure of eqs.(\ref{F_fourier}) is required by 
(anti--) self--duality. With eqs.(\ref{eta_eta}, \ref{F_fourier}) 
the fermion vertex resulting from the (anti--) instanton field 
strength, eq.(\ref{Y_1}), becomes
\be
(Y_{F\pm})^a_{\mu\nu}(x)[\psi^\dagger , \psi ]
&=& \frac{i M \bar\rho^2}{N_c} 
\fourint{k_1}\fourint{k_2} \; G(k) \exp (i k\cdot x )\;
F(k_1 ) F(k_2 ) \nonumber \\
&& \times \; \psi^\dagger_{i\alpha} (k_1 ) 
\left[ \Gamma_{\mu\nu} \frac{1 \pm \gamma_5}{2} \right]_j^i
\left(\frac{\lambda^a}{2}\right)^\alpha_\beta \psi^{j\beta} (k_2 ) ,
\nonumber \\
\Gamma_{\mu\nu} &=& \sigma_{\rho\nu} \frac{k_\rho k_\mu}{k^2}
+ \sigma_{\mu\rho} \frac{k_\rho k_\nu}{k^2}
- \half \sigma_{\mu\nu}, \hspace{1.5cm} 
k \; = \; k_2 - k_1 . \label{Y_2} 
\ee
With $F_{\mu\nu}^a$ replaced by these non-local quark vertices
resulting from instantons and antiinstantons, 
the operator $O_\sigma (x)$ is thus represented by the effective quark 
operator
\be
\effop{O_\sigma}(x)[\psi^\dagger , \psi ] &=& 
i \psi^\dagger (x) \frac{\lambda^a}{2} \sigma_{\mu\nu}\, 
\psi (x) \left[ (Y_{F+})^a_{\mu\nu}(x)
\; + \; (Y_{F-})^a_{\mu\nu}(x) \right] .
\label{effop}
\ee
A graphical representation of the effective operator
is given in fig.1a and b.
\par
We now calculate the average of eq.(\ref{effop}) in the vacuum of the 
effective quark theory derived from the instanton medium. 
The quark propagator of the effective theory includes the dynamically 
generated quark mass, $S(k) = (\kslash - i M F^2 (k))^{-1}$. Due to the 
color structure of eq.(\ref{effop}), the only non-vanishing contribution 
to the VEV is given by the diagram of fig.1c,
\be
\langle \effop{O_\sigma} (0) \rangle_{\rm eff} &=& 
\frac{N_c}{2\pi^2} M \bar\rho^{-4} \, I^{(2)}(M\bar\rho ) ,
\label{O_sigma_eff_aver}\\
I^{(2)}(M \bar\rho ) &\equiv&
4 \pi^2 \bar\rho^6 \fourint{k_1}\fourint{k_2} \; 
\frac{G(k) \, F(k_1 ) \, F(k_2 ) N(k_1 , k_2 )}
{\left[ k_1^2 + M^2 F^4 (k_1) \right] 
\left[ k_2^2 + M^2 F^4 (k_2) \right]} ,
\label{I2} \\
N(k_1 , k_2 ) &=& \textfrac{1}{4} 
\tr [\sigma_{\mu\nu} (\kslash_1 + i M(k_1 ))
\Gamma_{\mu\nu} (\kslash_2 + i M(k_2 )) ] \nonumber \\
&=& \frac{1}{k^2} \left[ 8 k_1^2 k_2^2 - 6 (k_1^2 + k_2^2 )
k_1\cdot k_2 + 4 (k_1\cdot k_2 )^2 \right] . \nonumber
\ee
The two--loop integral, eq.(\ref{I2}), is UV--convergent, since
$G(k) \sim k^{-4}$ and $F(k) \sim k^{-3}$ for $k\rightarrow\infty$.
It is also IR--finite, since the singularity of $G(k)$ 
at $k = 0$ is only logarithmic. Note that $I^{(2)}$ 
is dimensionless and depends only on the 
dimensionless product $M\bar\rho$. Numerical evaluation gives 
$I^{(2)}(M\bar\rho = 0) = -4$. (The parametric smallness 
of $M^2\bar\rho^2$, which is a consequence of the diluteness
of the instanton medium, is always implied when working
with the effective quark theory.
With the phenomenological value, $M\bar\rho \approx 0.6$, one obtains 
$I^{(2)} = -3.9$, which differs from the $M\bar\rho \rightarrow 0$ 
limit by only $3 \% $.) 
Thus, with the standard parameters of the instanton vacuum
\cite{DP86}, $\bar\rho^{-1} = 600\,{\rm MeV},\, M = 345\, {\rm MeV}$, 
we obtain a value of
\be
\langle \effop{O_\sigma} (0) \rangle_{\rm eff} &=& - (490\,{\rm MeV})^5 .
\label{result_GeV}
\ee
Within our scheme of approximations this is the result for the mixed 
condensate in QCD.
\par
The above calculation can easily be generalized to an arbitrary number
of quark flavors, $N_f > 1$. In this case, the effective fermion vertex
corresponding to the gluon operator $F_{\mu\nu}^a$
is a $2N_f$--fermion vertex. It has the form of eq.(\ref{Y_2})
for one flavor $f$ times a 't Hooft determinant of the remaining flavors, 
$f' \neq f$, summed over all flavors $f$ \cite{DPW95}. To leading order
in $1/N_c$ the vacuum averages factorize in eq.(\ref{O_sigma_eff_aver})
times the vacuum average of the 't Hooft determinant, which is unity
by virtue of the self--consistency condition determining the dynamical
quark mass, $M$ \cite{DP86,DPW95}. Thus, for $N_f > 1$ flavors the mixed 
condensate per flavor is again given by eqs.(\ref{O_sigma_eff_aver}, 
\ref{result_GeV}).
\par
Let us compute the ratio of the mixed condensate to 
the quark condensate directly within the instanton vacuum.
The quark condensate is given by the trace of the massive 
quark propagator of the effective quark theory \cite{DP86},
\be
-i \langle \psi^\dagger \psi \rangle_{\rm eff} &=&
- \frac{N_c}{2\pi^2} M \bar\rho^{-2} I^{(1)}(M\bar\rho ) , 
\label{quark_cond} \\
I^{(1)}(M\bar\rho ) &=& 8 \pi^2 \bar\rho^2 \fourint{k} 
\frac{F^2(k)}{k^2 + M^2 F^4 (k)} ,
\ee
where $I^{(1)}(M\bar\rho  = 0) = 1$. For the parameter $m_0^2$ of 
eq.(\ref{m_0_def}) we thus obtain
\be
m_0^2 &=& - \bar\rho^{-2} \; 
\frac{I^{(2)}(M \bar\rho )}{I^{(1)}(M \bar\rho )} .
\label{m_0_integrals}
\ee
This quantity is positive, since $I^{(1)}(M\bar\rho ) > 0$ and 
$I^{(2)}(M\bar\rho ) < 0$ for all realistic values of $M\bar\rho$.
Note that the explicit factor of $M$ in both eq.(\ref{O_sigma_eff_aver})
and eq.(\ref{quark_cond}) has canceled, so that $m_0^2$ depends on the
dynamical quark mass only through the integrals
$I^{(1)}(M\bar\rho ),\, I^{(2)}(M\bar\rho)$. This dependence is very 
weak. In particular, $m_0^2$ has a 
finite limit for $M\bar\rho \rightarrow 0$,
\be
m_0^2 &=& 4 \bar\rho^{-2} \hspace{2cm} (M\bar\rho \rightarrow 0).
\label{m_0_limit}
\ee
With the standard value of $\bar\rho$ this comes to
$m_0^2 = 1.4\,{\rm GeV}^2$.  
(At $M\bar\rho = 0.6$ one obtains $m_0^2 = 4.5\,\bar\rho^{-2}$.)
\par
The limiting value of $m_0^2$ for $M\bar\rho \rightarrow 0$
can be understood in simple terms. The instanton medium is 
characterized by two parameters, the instanton density, $N/V$, and 
average size, $\bar\rho$. Parametrically, the dynamical quark mass
behaves as $M \propto (N/V)^{1/2} \bar\rho$, so that 
the limit $M\bar\rho\rightarrow 0$ corresponds to the dilute 
limit of the instanton medium, $(N/V) \bar\rho^4 \rightarrow 0$. 
If $m_0^2$ approaches a finite value in this limit,
it must be possible to explain the limiting value considering only 
one single (anti--) instanton and its fermionic zero mode. 
Indeed, with the concrete form of the zero mode wave 
function in $x$--space, 
$\Phi_\pm^{i\alpha} (x)$, given {\em e.g.}\ in \cite{VZNS82,DP86}, 
one finds that the integral $I^{(2)} (M\bar\rho = 0)$ is nothing but
the ``matrix element'' of $O_\sigma$ (with $F_{\mu\nu}^a (x)$ being 
the instanton field) between zero mode wave functions,
\be
\langle \Phi_\pm | \half \lambda^a \sigma_{\mu\nu} F_{\pm\mu\nu}^a | 
\Phi_\pm \rangle &\equiv& \int d^4 x \,
\Phi_{\pm i\alpha}^\ast (x) \half (\lambda^a )^\alpha_\beta 
(\sigma_{\mu\nu})^i_j F_{\pm\mu\nu}^a
(x) \Phi_\pm^{j\beta} (x) 
\nonumber \\
&=& \bar\rho^{-2} I^{(2)} (M\bar\rho = 0) .
\ee
Similarly, one may verify that $I^{(1)} (M\bar\rho = 0)$ is equal 
to the normalization integral of the zero mode wave function,
\be
\langle \Phi_\pm | \Phi_\pm \rangle &\equiv& \int d^4 x \,
\Phi_{\pm i\alpha}^\ast (x) \Phi_\pm^{i\alpha} (x)
\;\; = \;\; I^{(1)} (M\bar\rho = 0) .
\ee
Thus, in the dilute limit, $M\bar\rho = 0$, eq.(\ref{m_0_integrals})
can be rewritten as
\be
m_0^2
&=& - \frac{\langle \Phi_\pm | \half \lambda^a 
\sigma_{\mu\nu} F_{\pm\mu\nu}^a | \Phi_\pm \rangle}
{\langle \Phi_\pm | \Phi_\pm \rangle} 
\label{m_0_single_inst}
\ee 
Expressing the instanton field as
\be
\half \lambda^a F_{\mu\nu}^a (x) &=& i [\nabla_\mu , \nabla_\nu ],
\hspace{2cm} \nabla_\mu \;\equiv\; \partial_\mu - i \half \lambda^a 
A_\mu^a (x) ,
\ee
and making use of the zero mode Dirac equation, 
$\nablaslash \Phi_\pm = 0$, one has
\be
\langle \Phi_\pm | \half \lambda^a \sigma_{\mu\nu} 
F_{\pm\mu\nu}^a | \Phi_\pm \rangle
&=& 2 \langle \Phi_\pm | \nabla^2 | \Phi_\pm \rangle .
\ee
A brief calculation, using elementary instanton algebra, then
leads from eq.(\ref{m_0_single_inst}) to eq.(\ref{m_0_limit}).
We note that this instanton result gives a precise meaning
to a qualitative argument of \cite{RRY85} relating the
mixed condensate to the ``average squared momentum'' of the vacuum 
quarks.
\par
Although eq.(\ref{m_0_single_inst}) seems intuitively obvious, we should 
keep in mind that its origin is non-trivial: 
both the quark condensate and the mixed condensate vanish in 
the limit $M\rightarrow 0$ and thus can not be obtained from
a single instanton. However, given the existence of chiral symmetry 
breaking in the instanton medium at finite density, the ratio of the 
two condensates is non-zero in the dilute limit and can be explained 
as a one--instanton property --- the ratio of the zero--mode matrix 
elements of the corresponding operators. 
\par
When comparing our results with the value of the mixed condensate 
extracted from QCD sum rules \cite{Ioffe81} we have to take into account 
the dependence on the normalization point. To lowest order, the
scale dependence of $\langle O_\sigma \rangle$ is given by
\be
\langle O_\sigma \rangle (\mu ) &=& 
\left(\frac{\alpha_s (\mu_{\rm inst} )}{\alpha_s (\mu )} 
\right)^{\gamma_\sigma /b}
\langle O_\sigma \rangle (\mu_{\rm inst}) , 
\label{evolution}
\ee
where $b = 11 N_c /3 - 2 N_f /3$ and $\alpha_s (\mu )$ is the
one--loop running coupling constant. Here, $\mu_{\rm inst}$ denotes the
scale at which the coupling constant is defined in the instanton
vacuum. (We do not need to consider mixing of
$O_\sigma$ with the $d = 6$ four--quark condensates here, since we are 
working in the chiral limit of vanishing current quark mass.)
The anomalous dimensions of $O_\sigma$
operator is small, $\gamma_\sigma = -2/3$ \cite{VZS76}. 
When expressing our result in the form eq.(\ref{result_GeV}) we can 
therefore neglect the scale dependence; this approximation is in fact
used in most of the applications of QCD sum 
rules \cite{SVZ79,Ioffe81,BI82}.  
\par
When considering the ratio of the mixed condensate to the quark
condensate, however, the scale dependence becomes essential:
the effective anomalous dimension of the parameter $m_0^2$ 
is $\gamma_\sigma - \gamma_{\bar\psi\psi} = -14/3$. 
The value of $m_0^2$ thus strongly decreases with
increasing normalization point. This makes it difficult to compare 
values of $m_0^2$ obtained in different approaches, the more since
the QCD sum rule estimates of \cite{BI82} are based on a value of 
$\Lambda_{\rm QCD} = 150\, {\rm MeV}$, which is too small in the
light of modern experimental results for $\alpha_s$.
(For a discussion of the effect of the value of $\Lambda_{\rm QCD}$ on 
QCD sum rules, see \cite{Shifman95}.)
Moreover, at the level of approximations considered here, 
it is not possible to precisely pin down the normalization point of 
the instanton result, eq.(\ref{m_0_limit}). The scale 
$\mu_{\rm inst}$ is of order $\bar\rho^{-1} = 600\,{\rm MeV} $, but 
there may be a factor of order unity.
Nevertheless, comparing eq.(\ref{m_0_limit}) with
the QCD sum rule estimate of \cite{BI82}, 
$m_0^2 = 0.8 \pm 0.2\, {\rm GeV}^2$ at 
$\mu = 500\,{\rm MeV}$, identifying for a moment the normalization
points, it seems that that our result is larger by $\sim 50 \%$.
We stress again that the ambiguity in the normalization point
concerns only the representation of the mixed condensate
through the parameter $m_0^2$, eq.(\ref{m_0_def}), not the
value of the mixed condensate itself, eq.(\ref{result_GeV}).
\par
To summarize, we have shown that the instanton vacuum naturally
leads to a mixed quark--gluon condensate whose sign and order of
magnitude agree with the QCD sum rule estimate. The ratio of the
mixed condensate to the quark condensate, $m_0^2$, allows a simple
interpretation as the ratio of matrix elements between zero--mode
wave functions of one instanton. The relatively large value of
this quantity can be seen as a consequence of the smallness
of instantons.
\par
The effective operator formulation of \cite{DPW95} provides a simple
and systematic method to evaluate matrix elements of gluon operators
such as eq.(\ref{O_sigma}). It allows not only the computation
of vacuum condensates, but also of nucleon matrix 
elements of gluon operators which arise in the description of
power corrections in deep--inelastic scattering. Work in this 
direction is in progress. 
\\[1cm]
We thank D.I.\ Diakonov for his advice and interest in 
this investigation, as well as V.Yu.\ Petrov and P.V.\ Pobylitsa for many 
helpful conversations.
\newpage
\begin{figure}[14.5cm,14.5cm]
\vspace{-1cm}
\epsfxsize=14.5cm
\epsfysize=14.5cm
\centerline{\epsffile{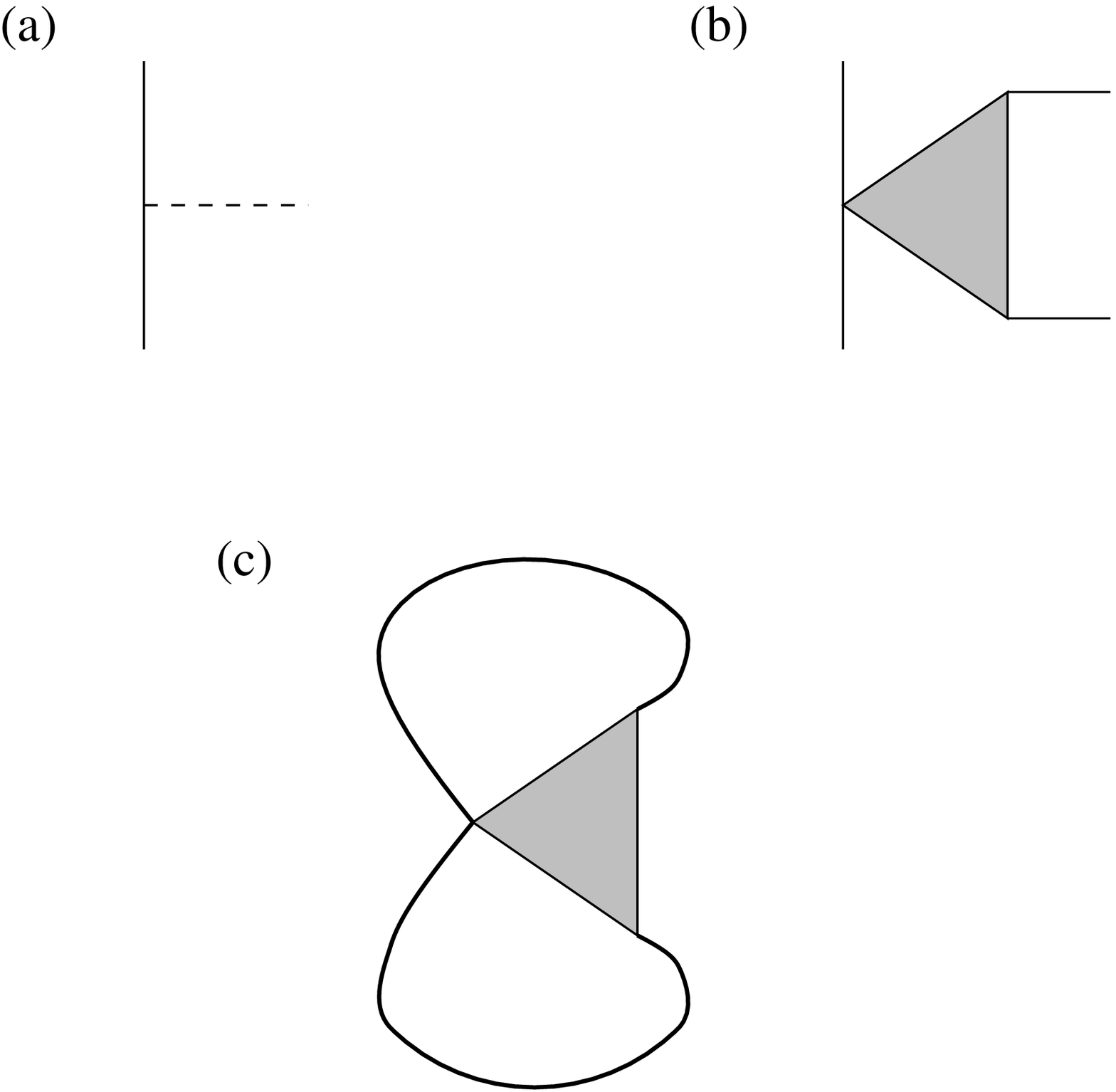}}
\vspace{1cm}
\caption[]{A graphical representation of the effective operator
approach. (a) The QCD operator, $O_\sigma$, eq.(\ref{O_sigma}).
The dashed end denotes the gluon field.
(b) The effective many--fermion operator, 
$\effop{O_\sigma} [\psi^\dagger , \psi ]$, 
eq.(\ref{effop}). The gluon field is replaced by the non-local 
fermion vertices, $Y_{F\pm}$. Shown is the case 
of one quark flavor, $N_f = 1$; in general, $Y_{F\pm}$ is a 
$2 N_f$--fermion vertex. (c) The VEV of the effective operator in the
effective quark theory, eqs.(\ref{O_sigma_eff_aver}, \ref{I2}). 
The fat lines denote the massive quark 
propagator, $S(k) = (\kslash - i M F^2 (k))^{-1}$.}
\end{figure}
\end{document}